# How to prolong network life-span in wireless networks

*Dana Anderson*
*University of Colorado at Boulder, CO30238, USA*
Email: anderson2dana@hotmail.com

**Abstract:** One of the most important problems in wireless sensor network is to develop a routing protocol that has energy efficiency. Since the power of the sensor Nodes are limited, conserving energy and network life is a critical issue in wireless sensor network. Clustering is one of the known methods widely used to face these challenges. In this paper, a cluster based communication protocol with considering the low energy consumption in wireless sensor networks, is introduced which balances the energy load among sensor nodes. The nodes close to each other have more overlap; they sense the same data from environment and cause a waste of energy by generating repetitive data. In this paper, a cluster based routing protocol is introduced, in the proposed protocol, in each round a certain number of nodes are specified; the nodes which have at least one neighboring node at a distance less than the threshold. Then, among them the nodes with less energy and greater overlap with their neighbors have been chosen to go to sleep mode, Also, the energy imbalance among sensor nodes is reduced by integrating the distance of the nodes from the base station into clustering policies. A simulator was developed in the MATLAB environment to evaluate the performance of this protocol. On the basis of the simulation results the proposed protocol increases the network lifetime, balances the load among sensor nodes further, and increases the network coverage.

## INTRODUCTION

With advances in processor and wireless communication technologies, WSN will be used everywhere in the future life. It is a type of wireless ad-hoc networks and composed of a collection of sensor nodes. Within a WSN, nodes collaborate amongst themselves to accomplish a common task. WSN has enormous potential to improve how we use information from the real world. It is expected to apply the wireless sensor network technology to various application areas such as the military, environment, medical service, and business. Among the various scopes, one of the major applications of sensor network is to collect information periodically from a remote terrain where each node continually sense the environment and sends back this data to the base station (BS) which is usually located at Considerably far from the target field [1].

In most of the applications sensors are required to detect events and then communicate the collected information to a distant base station (BS) where parameters characterizing these events are estimated. The cost of transmitting information is higher than computation and hence it is be advantageous to organize the sensors into clusters [2] [3], where the data gathered by the sensors is communicated to the BS through a hierarchy of cluster-heads.

Sensor nodes are usually battery-powered and operate without attendance for a relatively long period of time. Therefore, energy efficiency becomes critical importance. In the wireless sensor network community, a significant focus has been on put on increasing energy efficiency. In recent years, one of the major researches in wireless sensor network is energy efficient routing protocol. Routing protocols for sensor network are very much application dependent [4].

The nodes close to each other have more overlap; they sense the same data from environment and cause a waste of energy by generating repetitive data. Therefore, in the proposed protocol, in each round a certain number of nodes, having at least one neighboring node at a distance less than the threshold, are specified. Then, among them the nodes with less energy and greater overlap with their neighbors are chosen to go to sleep mode, Also, in each round the nodes having more energy are selected as cluster head. In this paper, a new routing protocol, based on clustering for wireless sensor networks, is presented. Main objectives of this method are network load balancing, and increase in longevity and coverage of the network.

One important issue in the design of wireless sensor networks is to balance the energy load among the sensors in the network. However, current hierarchical protocols (LEACH [2], [3], TEEN [5], APTEEN [6]) do not consider the effect of distance parameter on sensors energy dissipation. In this paper, a cluster based protocol has developed that significantly decreases the energy imbalance in the network by integrating the distance of the sensor nodes from the BS into clustering policies. Furthermore, the proposed protocol does not require any centralized support From the BS and each sensor node makes its decision about whether to be a cluster head independently and thus this scheme offers a scalable routing protocol [7].

The rest of the paper is organized as follows: Section II gives a short overview of some related work concerning the proposed protocol while section III describes briefly the network coverage analysis, section IV presents our proposed protocol. Simulation results of our protocol are presented in section V. Finally, section VI concludes this paper.

## RELATED WORKS

Low-Energy Adaptive Clustering Hierarchy (LEACH) presented in [2], [3], provides an elegant hierarchical protocol that uses localized coordination to enable scalability and robustness for sensor networks, and exploits data aggregation in the routing protocol to reduce the amount of data packets that must be transmitted to the BS. LEACH divides the operation of the entire network into many rounds. Each round consists of a set-up phase and some number of time frames that construct the steady-state phase. During the set-up phase some sensor nodes elect themselves as cluster heads and announce their cluster head position to the rest of the nodes in the network, and then other nodes organize themselves into local clusters by choosing the most appropriate cluster head normally the closest cluster head. During the steady-state phase the cluster heads receive sensor data from cluster members, and transfer the aggregated data to the BS.

In LEACH, the decision of each node to become cluster head is taken based on the suggested percentage of cluster head nodes $p$ (programmed into the nodes prior to network operation), which is equal to $k_{opt}/N$, current round $r$, and an indicator function that shows whether or not the node has been cluster head in recent ($r \mod (1/p)$) rounds denoted by $C_i(t)$. Each sensor $i$ may become cluster head at the beginning of round $r$ (which starts at time $t$) with probability:

$$p_i(t) = \begin{cases} \dfrac{p}{1 - p \times (r \mod \dfrac{1}{p})} & if \ C_i(t) = 1 \\ 0 & if \ C_i(t) = 0 \end{cases} \quad (1)$$

Note that $C_i(t)$ is one if node $i$ is eligible to be a cluster head at time $t$ and zero otherwise. Therefore, every node will be a cluster head at some point within $N/k = 1/p$ rounds.

## THE NETWORK COVERAGE ANALYSIS

This section presents a method to analyze the network coverage [8]. The sensors are deployed in a two dimensional field with a certain density (number of nodes per unit area) where Boolean sensing model is adopted. Let $q$ be a random chosen point in the sensor field. Considering no scheduling algorithm is applied (all sensors are on) this point is covered when there is at least one sensor in the circle A of radius $r$ around $q$. This circle has area $\pi r^2$ and the probability to find at least one sensor is:

$$f_a = 1 - e^{-\lambda \pi r^2} \quad (2)$$

This equation is only used when all nodes are on. However, in this protocol whenever it is necessary a set of sensors are off (sleep) and the rest of them are on (active). So to involve the active nodes, equation (2) is modified by adding a new term representing the average number of active rounds for each node, which expresses the suggested coverage probability as equation (3).

$$f_a = 1 - e^{-\lambda \pi r^2 (\frac{t_a}{T})} \quad (3)$$

Where $\lambda$ is the network density, $r$ is the sensing range, $t_a$ is average active rounds of the node, and $T$ is the total number of rounds. Based on equation (3), if it is required to achieve the network coverage equal to a certain $f_a$, the required low bound of the network density to fulfill the required coverage can be computed by solving the equation (4).

$$\lambda_{(f_a)} = -\dfrac{\log(1 - f_a)}{\pi r^2 \left(\dfrac{t_a}{T}\right)} \quad (4)$$

Based on the simulation results, the value of ($t_a/T$) is almost 0.53 for different number of sensor nodes deployed (in the range between 200 to 1200). As an example, if the required coverage is equal to 0.9, sensing range is $r = 10$ meters, and the area is 100 meters by 100, the required network density equals to 0.0138, which means that 138 sensors nodes are required in the whole network to achieve a coverage of 0.9.

## PROPOSED PROTOCOL

The power of nodes in wireless sensor networks is limited, so one of the most important targets in wireless sensor networks is to increase the network lifetime and to distribute the load uniformly across the network.

In wireless sensor networks, the less distance with each other the nodes have, the more overlap they have, and nods covering almost the same area receive (or sense) the same data from the environment which consequently causes to waste energy as the following:
1. Sensing the same data.

2. Repetitive Data aggregation in cluster head
3. Sending repetitive data to base station

The solution proposed to handle this problem in the presented protocol is to make the nodes to go to sleep mode. In order to choose the nodes which are to go to sleep we heed three factors: 1. the energy level, 2. the number of neighbors, 3. the average distance among neighbors. The more the number of the neighbors of a node is and the less the average distance from its neighbors is, the more overlap it has. Moreover, the nodes having the lower energy levels are more likely to die. That being the case, the network lifetime as well as the network coverage decreases more.

The proposed protocol, like LEACH protocol, divides time, in parts with equal length which are called ROUND. Each ROUND is divided into two phases. The first one called set-up phase is a phase for cluster formation and cluster head selection. The second one called the steady-state phase is a phase related to normal network performance and data transfers.

In set-up phase, firstly, we find pair nodes which their distance is less than $d_{max}$ and we choose, between these two nodes, the node with lower energy level as the one which has the condition to go to sleep mode. Then, in order that there appear no holes in network coverage and network connection does not decrease, we ensure that when one of these pair nodes goes to sleep mode, the other one remains in active mode [9], Moreover it is assumed that nodes are aware of their approximate distance from each others.

After finding all nodes that have the conditions to go to sleep, if the number of these nodes is less than the maximum number of nodes which can go to sleep (*maxsleep*), we put all nodes into sleep mode; otherwise we sort the descending nodes according to their *NTE* amount, and put them-the *maxsleep* ones-in sleep mode. Afterward the system goes into the steady state phase. The Way to reach this factor (*NTE*) is described in equation (5):

$$NTE = \frac{N}{E^2 AVE} \quad (5)$$

In equation (5), *AVE* is the average distance of the node from its neighboring nodes, *N* is the number of neighboring nodes, and *E* is the energy level of the nodes. As it is obvious in equation (5), the node with lower energy level (*E* value is always smaller than 1), with higher number of neighboring nodes, and with lower average distance to its neighboring nodes has higher *NTE* amount.

The clustering paradigm increases the burden on the cluster heads, forcing them to deplete their batteries much faster than non-cluster head nodes. Therefore, a potential problem with LEACH protocol is that it assigns the probabilities of being cluster head to each node, assuming that being a cluster head consumes approximately the same amount of energy for each node. While, according to the radio model in the section V, the minimum required amplifier energy is proportional to the square of the distance from the transmitter to the intended Receiver [10]. As a result, the energy required by a far cluster head to communicate with the BS may be so large that the near cluster head can communicate four times (and even more) with the BS using the same energy. Therefore after the network operates for some rounds, there will be noticeable difference on energy consumption between the nodes near the BS and those far from the BS. If all nodes begin at the same energy level, the nodes far from the BS will exhaust their energy before those near the BS. As a result, the network will be partitioned into dead-areas and live-areas and hence the performance of the network will decline [7].

Proposed protocol divides the sensor field into multiple segments with equal area by drawing concentric circles around the BS and differentiates the number of clusters for each segment in terms of the distance from the BS. In closer segments the probability of becoming cluster head is more than distant segments and thus the number of cluster heads in these segments is more [7].

For the development of the proposed protocol, the same assumptions as in [2], [3], are made about the sensor network. Moreover it is assumed that nodes are aware of their approximate distance from the BS, so that the sensor nodes can guess the segment they belong. In [11], the authors have used signal strength parameter for approximating the distance parameter [7].

As stated before, the same cluster formation sequence as LEACH is used in proposed protocol, but the cluster head selection method [7] and the method of selecting nodes to go to sleep mode are different from LEACH.

This method increases the network lifetime for three reasons:
1. The nodes in sleep mode do not consume energy. It is generally to the benefit of saving energy and increasing network lifetime.
2. Less energy is consumed by cluster heads to aggregate repetitive data.
3. The nodes near to BS become more cluster head than the ones further away from it, therefore they consume less energy to communicate with the BS.

Putting some of the nodes into sleep mode reduces network coverage. So we tried to select nodes which have a lot of overlap with their neighbors. As it is described in section 3, we can calculate the amount of *maxsleep* for a certain percentage of network coverage; however, coverage is always greater than this value, because nodes going to sleep mode have at least one neighboring node with a distance less than $d_{max}$. The Neighboring node prevents the reduction of network coverage up to the amount of overlap it has with the node going to sleep.

## SIMULATION RESULTS

In this section the performance of the proposed protocol has been evaluated by using various tests. MATLAB is used to perform simulation. In this simulation, the performance of the proposed protocol has been compared with leach protocol, and the amount of radio range and sensing range of sensor nodes is considered 10 meters. Parameters used in the simulation are given in Table I.

TABLE 1. parameters values used in the simulation

| parameter | value |
|---|---|
| Network range | (0,0) to (100,100) |
| Number of nodes | 150 |
| $d_0$ | 87.7 m |
| Base station position | (50,50) |
| $E_{elec}$ | 50 nJ/bit |
| $\varepsilon_{fs}$ | 10 pJ/bit/m$^2$ |
| $\varepsilon_{mp}$ | 0.0013 pJ/bit/m$^4$ |
| Maximum number of segments | 10 |
| $P_{LEACH}$ | 0.1 |
| Number of time frames in each round | 1 |
| $E_{DA}$ | 5 nJ/bit/signal |
| The initial energy of each node | 0.1 J |
| Packet size | 500 bytes |

Probabilities of being a cluster head for nodes of each segment are given in Table II. In closer segments the probability of becoming cluster head is more than distant segments.

TABLE 2. Probabilities of being a cluster head for nodes of each segment

| Segment number | Probability of being a cluster head |
|---|---|
| 1 | 0.1 |
| 2 | 0.95 |
| 3 | 0.9 |
| 4 | 0.85 |
| 5 | 0.8 |
| 6 | 0.75 |
| 7 | 0.7 |
| 8 | 0.65 |
| 9 | 0.6 |
| 10 | 0.55 |

We use the same radio model as stated in [2], [3], [10], with $E_{elec}$ as energy being dissipated to run the transmitter or receiver circuitry and $\varepsilon_{amp}$ as the energy dissipation of the transmission amplifier. Transmission ($E_{Tx}$) and receiving ($E_{Rx}$) costs are calculated as equation (6):

$$E_{Tx}(l,d) = lE_{elec} + l\varepsilon_{amp}d^n$$
$$E_{Rx}(l,d) = lE_{elec} \qquad (6)$$

with $l$ as the length of the transmitted/received message in bits, $d$ as the distance between transmitter and receiver node, and $n$ as the path loss exponent which is two for the free space model and can be up to six depending on the environment and network topology [10]. As it can be seen, the transmitter expends energy to run the radio electronics and power amplifier, while the receiver only expends energy to run the radio electronics. In this paper, the free space model ($n = 2$, $\varepsilon_{amp} = \varepsilon_{fs}$) is used for the transmission distances below a threshold distance $d0$ with typical value of 87.7 m, and multipath model ($n = 4$, $\varepsilon_{amp} = \varepsilon_{mp}$) is used for further distances.

In cluster-based schemes, the cluster heads are responsible for aggregating the data signals of their cluster members to produce a single representative signal, expending $lE_{DA}$ for each $l$-bit input signal, where $E_{DA}$ is the energy for data aggregation and is set to 5 nJ/bit/signal.

441 Hypothetical points in simulation environment (100 meters*100meters) have been used to compare network coverage of the proposed protocol with LEACH protocol. And the amount of the coverage on these points is measured in each round of the protocol. The point locations are shown in Figure 1.

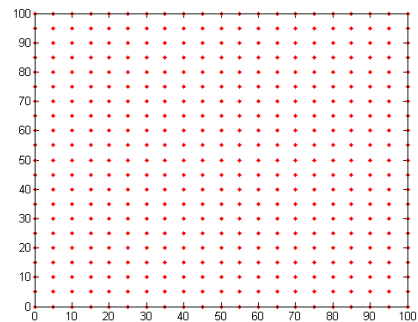

Fig. 1 the point locations.

As it is mentioned in Section 3, the required number of nodes to achieve a coverage of 0.9 in the network is at least 138. Since the total number of nodes is 150 the number of nodes which can be put into sleep mode without a coverage less than 0.9 is 12. Therefore, the amount of *maxsleep* is 12. The amount of $d_{max}$ is 3.5 meter because for amounts less than 3.5, the number of nodes that could go to sleep mode always less than *maxsleep*.

Figure 2 shows the total number of nodes that remain alive over simulation time of 800 rounds. It can be seen that nodes remain alive for a longer time in the proposed protocol than LEACH.

Using two metrics, First Node Dies (FND) and Half of the Nodes Alive (HNA) proposed in [12], we exactly compare LEACH with the proposed protocol in terms of network lifetime. Figure 3 shows that using the proposed protocol can increase the lifetime of a sensor network by 16% for FND and more than 16% for HNA.

Figure 3 shows the total dissipated energy over simulation time of 800 rounds. It can be seen that the proposed protocol consumes lower energy than LEACH.

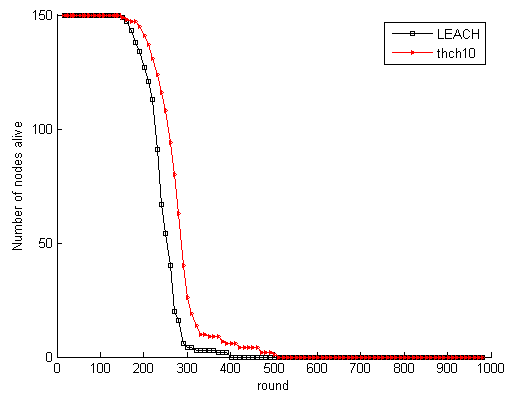

Fig. 2 System lifetime using LEACH, proposed protocol.

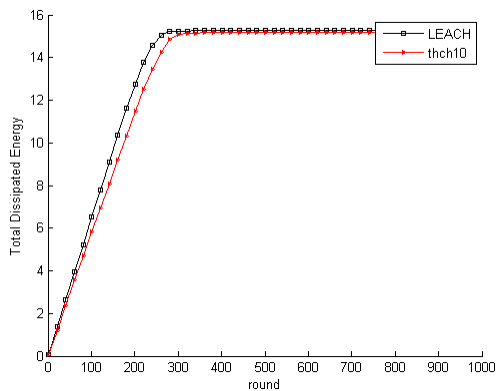

Fig. 3 total dissipated energy using LEACH, proposed protocol

Figure 4 shows the network coverage over simulation time of 800 rounds. Apparently the proposed protocol network coverage is higher than LEACH.

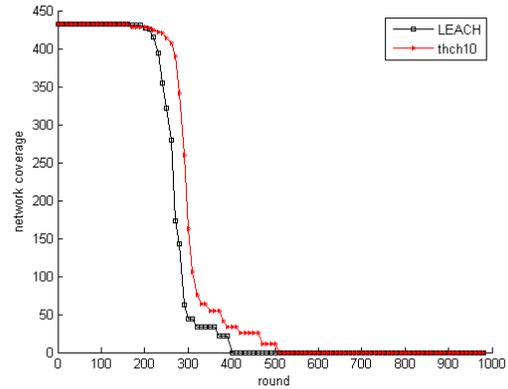

Fig. 4 network coverage using LEACH, proposed protocol

Figure 5 shows the Variance of energy level of nodes over simulation time of 800 rounds. The proposed protocol variance is less than LEACH, which shows: the proposed protocol has better load balancing than LEACH protocol.

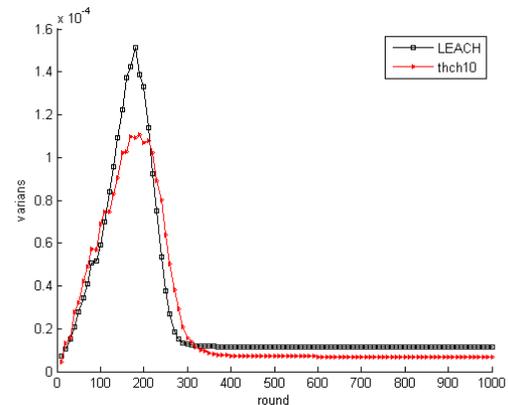

Fig. 5 Variance of energy level of nodes using LEACH, proposed protocol

## CONCLUSION

In this new approach some of problems of previously proposed LEACH algorithm are eliminated. According to the simulation results, compared to Leach protocol, it is possible to decrease the power consumption in proposed protocol, increase network lifetime and network coverage by means of putting some nodes into sleep mode.

In each round of the implementation of this protocol, it has been tried to select nodes having more overlap with their neighbors and lower energy to put in to sleep mode. These nodes have mostly sense similar data with the neighbors and cause waste of energy in the following operations: sensing data, sending the data, and data aggregation. The huge difference between energy levels of near nodes and far nodes has been compensated by dividing the sensor field into some segments and applying different probabilities to them.

This method can prevent wasting of energy and consequently cause to increase the network lifetime.

The proposed protocol can be improved by selecting the nodes with more energy for cluster head, using multi-hop routing between cluster head and base station and applying different initial energy levels according to the distances of the nodes from the BS.